\pdfoutput=1
\documentclass[sigconf,screen]{acmart}
\setcopyright{cc}
\setcctype{by-nc-nd}
\acmDOI{10.1145/3832783.3837437}
\acmYear{2026}
\copyrightyear{2026}
\acmISBN{979-8-4007-2882-2/2026/10}
\acmConference[ASE '26]{Proceedings of the 41st IEEE/ACM International Conference on Automated Software Engineering}{October 12--16, 2026}{Munich, Germany}
\acmBooktitle{Proceedings of the 41st IEEE/ACM International Conference on Automated Software Engineering (ASE '26), October 12--16, 2026, Munich, Germany}
\acmSubmissionID{ase26main-p355-p}
\received{2026-03-26}
\received[accepted]{2026-06-18}

\usepackage{microtype}
\usepackage{subcaption}
\usepackage[ruled,vlined,linesnumbered]{algorithm2e}
\usepackage{graphicx}
\usepackage{tikz}
\usepackage{verbatim}
\usepackage{caption}
\usepackage{setspace}
\usepackage{arydshln}
\usepackage{multirow}
\usepackage{enumerate}
\usepackage{enumitem}
\usepackage{pifont}
\usepackage{listings}
\usepackage{color}
\usepackage[table]{xcolor}
\usepackage[most]{tcolorbox}
\usepackage{array}
\usepackage{booktabs}
\usepackage{threeparttable}
\usepackage{makecell}
\usepackage{amsmath}
\usepackage{xspace}
\usepackage{hyphenat}
\usepackage[normalem]{ulem}
\usepackage{balance}
\usetikzlibrary{arrows.meta, positioning, calc}

\newcommand{\ourmethod}{{{AgentChaos}}\xspace}
\newcommand{\llmname}[1]{{\texttt{#1}}\xspace}

\newcolumntype{C}[1]{>{\centering\arraybackslash}p{#1}}

\definecolor{mygrey}{gray}{0.4}

\definecolor{commentgray}{rgb}{0.4,0.4,0.4}
\definecolor{deepblue}{rgb}{0.1,0.1,0.6}
\definecolor{darkred}{rgb}{0.6,0.1,0.1}
\definecolor{bglight}{rgb}{0.98,0.98,0.98}
\lstdefinestyle{compressedStyle}{
    basicstyle=\ttfamily\scriptsize, 
    breaklines=true,
    xleftmargin=2pt,               
    xrightmargin=2pt,
    aboveskip=0pt,                 
    belowskip=0pt,                 
    frame=none,                    
}

\newtcolorbox{tightbox}[1]{
    colback=white,
    colframe=black!80,
    boxrule=0.4pt,               
    sharp corners,                 
    enhanced,
    breakable,
    boxsep=0pt,                     
    top=1pt,                      
    bottom=1pt,
    left=1pt,
    right=1pt,
    fonttitle=\bfseries\scriptsize,
    title=#1,
    attach title to upper,
    after title={\quad},         
}

\definecolor{dupcolor}{RGB}{255,200,100}
\lstdefinestyle{codestyle}{
    basicstyle=\ttfamily\scriptsize,
    frame=tb,
    columns=fullflexible,
    keepspaces=true,
    framexleftmargin=0.2em,
    framexrightmargin=0.2em,
}

\newcolumntype{C}[1]{>{\centering\arraybackslash}m{#1}}
\newcolumntype{L}[1]{>{\raggedright\arraybackslash}m{#1}}
\newcolumntype{R}[1]{>{\raggedleft\arraybackslash}p{#1}}

\definecolor{dkgreen}{rgb}{0,0.6,0}
\definecolor{gray}{rgb}{0.5,0.5,0.5}
\definecolor{mauve}{rgb}{0.58,0,0.82}
\definecolor{lightblue}{RGB}{245, 248, 254}
\definecolor{darkblue}{RGB}{134, 134, 215}

\newcommand{\fdbox}[1]{
\begin{tcolorbox}[tile, size=fbox, boxsep=2mm, boxrule=0pt, top=0pt, bottom=0pt,
borderline west={1mm}{0pt}{gray!50!white}, colback=gray!10!white]
#1
\end{tcolorbox}
}

\AtBeginDocument{%
  }

\begin{document}

\title{AgentChaos: Chaos Engineering for Agent Systems via Programmatic Fault Injection}

\author{Gou Tan}
\orcid{0009-0008-6580-1470}
\affiliation{%
  \institution{Sun Yat-sen University}
  \city{Guangzhou}
  \country{China}
}
\email{tang29@mail2.sysu.edu.cn}

\author{Zhensu Sun}
\orcid{0000-0001-5393-7858}
\affiliation{%
  \institution{Singapore Management University}
  \country{Singapore}
}
\email{zssun@smu.edu.sg}

\author{Jieke Shi}
\orcid{0000-0002-0799-5018}
\affiliation{%
  \institution{Singapore Management University}
  \country{Singapore}
}
\email{jiekeshi@smu.edu.sg}

\author{Ting Zhang}
\orcid{0000-0002-6001-1372}
\affiliation{%
  \institution{Monash University}
  \country{Australia}
}
\email{ting.zhang@monash.edu}

\author{Zilong He}
\orcid{0000-0001-7963-082X}
\affiliation{%
  \institution{Sun Yat-sen University}
  \city{Guangzhou}
  \country{China}
}
\email{hezlong@mail2.sysu.edu.cn}

\author{Qingfu Wu}
\orcid{0009-0008-9767-8366}
\affiliation{%
  \institution{Sun Yat-sen University}
  \city{Guangzhou}
  \country{China}
}
\email{wuqf23@mail2.sysu.edu.cn}

\author{Shuai Liang}
\orcid{0009-0008-0256-9037}
\affiliation{%
  \institution{Sun Yat-sen University}
  \city{Guangzhou}
  \country{China}
}
\affiliation{%
  \institution{China Unicom}
  \city{Beijing}
  \country{China}
}
\email{liangsh76@mail2.sysu.edu.cn}

\author{Weifeng Sun}
\orcid{0000-0001-6013-1369}
\affiliation{%
  \institution{Singapore Management University}
  \country{Singapore}
}
\email{wfsun@smu.edu.sg}

\author{Junda He}
\orcid{0000-0003-3370-8585}
\affiliation{%
  \institution{Singapore Management University}
  \country{Singapore}
}
\email{jundahe.2022@phdcs.smu.edu.sg}

\author{Pengfei Chen}
\authornotemark[2]
\makeatletter
\g@addto@macro\@authornotes{\footnotetext[2]{Pengfei Chen is the corresponding author.}}
\makeatother
\orcid{0000-0003-0972-6900}
\affiliation{%
  \institution{Sun Yat-sen University}
  \city{Guangzhou}
  \country{China}
}
\email{chenpf7@mail.sysu.edu.cn}

\author{Chuanfu Zhang}
\orcid{0009-0007-1525-5830}
\affiliation{%
  \institution{Sun Yat-sen University}
  \city{Guangzhou}
  \country{China}
}
\email{zhangchf9@mail.sysu.edu.cn}

\author{Lwin Khin Shar}
\orcid{0000-0001-5130-0407}
\affiliation{%
  \institution{Singapore Management University}
  \country{Singapore}
}
\email{lkshar@smu.edu.sg}

\author{David Lo}
\orcid{0000-0002-4367-7201}
\affiliation{%
  \institution{\makebox[0pt][c]{Singapore Management University}}
  \country{Singapore}
}
\email{davidlo@smu.edu.sg}

\renewcommand{\shortauthors}{Tan et al.}

\begin{abstract}
Agent systems rely on LLM APIs for every response, but these APIs can return server errors, truncated responses, or corrupted content that propagates through downstream agents and causes task failure. Evaluating robustness under these faults is crucial for reliable deployment. Existing fault injection methods are offline, require source code modification, or cannot modify specific response fields. A comprehensive evaluation also requires a systematic fault taxonomy because different fault types affect downstream agents differently. We propose AgentChaos, a chaos engineering framework for controlled, runtime, non-intrusive LLM API fault injection. Since all agent systems access LLMs through the same HTTP interface, we inject faults at this shared layer without modifying source code. We define crash, omission, and value faults on content and tool call fields, intercept and modify LLM API responses at runtime, and verify whether each fault is triggered to filter untriggered tasks and avoid underestimating fault impact. Evaluations across agent systems, benchmarks, and backbone LLMs under 65 fault configurations show that all systems degrade under fault injection, with pass@1 dropping by up to 50 percentage points. The ranking is consistent across models, suggesting that robustness depends on system implementation rather than model capability. Existing fault diagnosis methods achieve below 53\% accuracy on fault type and below 56\% on fault step, leaving room for improvement. We further reveal practical findings for agent system developers.
\end{abstract}




\setcounter{footnote}{1}
\maketitle
\section{Introduction}\label{sec:introduction}

Agent systems built on Large Language Models (LLMs) have been widely adopted for complex tasks, such as question answering~\cite{24LiangMAD}, reasoning~\cite{24WangMMLUPro, 24WangMATH500}, and software engineering~\cite{24WuAutoGen, 25XuScalable, 24IslamMapCoder, 25HuEvoMAC, 26TanLIDL}. The majority of these systems, such as Claude Code, rely on LLM APIs to generate every response.
To fulfill a single task, an agent system typically issues multiple LLM API calls, making the reliability of the underlying APIs a critical factor for the entire system.

\begin{figure}[t]
\centering
\includegraphics[width=\linewidth]{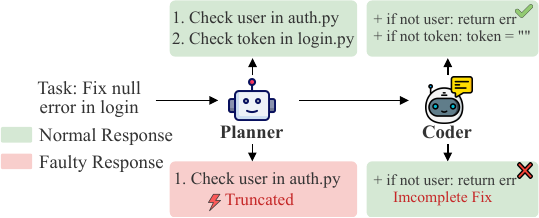}
\vspace{-14pt}
\caption{A truncated LLM API response at the Planner produces an incomplete plan, leading to an incomplete fix.}
\label{fig:01intro}
\vspace{-18pt}
\end{figure}

However, LLM APIs in practice can return various errors, such as server errors (e.g., HTTP 5xx), truncated responses (e.g., token limit cutoff), or corrupted content (e.g., garbled characters).
The more LLM API calls a task requires, the higher the chance of encountering such a fault.
As shown in Fig.~\ref{fig:01intro}, once a faulty response occurs, it can propagate through downstream agents and cause task failure.
Quantifying the impact of LLM API faults on agent systems is therefore crucial for reliable deployment. Chaos engineering provides a principled approach to this problem by injecting controlled faults into running systems and observing their behavior, discovering weaknesses before real failures occur in production~\cite{26YuFailure}.

Existing fault injection methods cannot support controlled, runtime, non-intrusive LLM API fault injection.
\textit{Agent-oriented fault injection methods} either operate offline or at runtime target non-API faults.
Offline methods (e.g., AgenTracer~\cite{25ZhangAgenTracer}) use an LLM to perturb completed execution traces after tasks finish, producing fault labels for downstream diagnosis rather than injecting faults into live systems. Because faults are never injected into running systems, they cannot trigger real runtime behaviors nor reveal actual software weaknesses.
Runtime methods (e.g., MAS-FIRE~\cite{26JiaMASFIRE}) target semantic failures such as hallucination and role ambiguity by hijacking system prompts, rewriting agent outputs, and rerouting messages. The injection logic needs to be re-implemented for each target system, and systems that do not expose the required interfaces cannot be injected.
Since these methods do not target LLM API faults, their findings on fault impact do not generalize to LLM API faults such as response truncation.
In contrast, \textit{traditional API fault injection methods} (e.g., Rainmaker~\cite{23ChenRainmaker}, ChaosBlade~\cite{26ChaosBlade}) operate at the HTTP or OS layer without code modification, but treat responses as opaque and cannot modify specific fields within the response body, such as \texttt{message.content} or \texttt{tool\_calls}. They can simulate server errors and timeouts, but cannot inject faults that require parsing and modifying the response content, such as response truncation or encoding corruption.

Beyond the injection method itself, a comprehensive robustness evaluation also requires a systematic fault taxonomy. LLM API faults take many forms, such as server errors, truncated responses, and corrupted encoding, and each form affects downstream agents differently. Without a systematic taxonomy, evaluators may test only a few obvious types and miss the most damaging ones. For example, a server error is easy to detect and retry, but a truncated response still looks like valid output, so the agent accepts it and passes the incomplete result to the next step.

To address these problems, we propose \ourmethod, a chaos engineering framework that evaluates agent system robustness through controlled, runtime, non-intrusive LLM API fault injection.
Our key insight is that all agent systems access LLMs through the same HTTP request-response interface, so we can inject faults at this shared transport layer without modifying any source code.
We first define a fault taxonomy by adapting the classical fault classification from distributed systems~\cite{04AvizienisBasic} to LLM API responses (\S\ref{sec:fault_classification}), covering crash, omission, and value faults on both content and tool call fields. Combined with injection strategies and compound scenarios, this taxonomy yields 65 fault configurations.
We then design two components for the injection framework. (1)~An \textit{HTTP-layer injection mechanism} (\S\ref{sec:method_injection}) that patches the HTTP client at runtime to intercept and modify LLM API responses according to the fault configuration, requiring no changes to any agent system. (2)~A \textit{trigger verification procedure} (\S\ref{sec:method_verification}) that records execution traces and checks whether each fault was actually triggered, filtering untriggered tasks from evaluation.
During evaluation, each task receives one fault configuration at runtime, and we compare pass@1 before and after injection to quantify robustness degradation.


We evaluate across multiple agent systems~\cite{24WuAutoGen, 24LiangMAD, 24IslamMapCoder, 25HuEvoMAC, 25XuScalable}, benchmarks~\cite{21ChenHumanEval, 23LiuHumanEvalPlus, 21AustinMBPP, 24WangMMLUPro, 24WangMATH500, 25DengSWEBenchPro}, and backbone LLMs~\cite{25Claude45, 25CHATGPT, 25DeepSeekV32, 25Seed} under all 65 fault configurations. All systems show performance degradation, with pass@1 dropping up to 50 percentage points (MapCoder on HumanEval+), as shown in Table~\ref{tab:overall}.
We also evaluate fault diagnosis on failed tasks, where the goal is to identify which type of fault was injected and at which LLM call the fault occurred. Both rule-based pattern matching and LLM-based diagnosis~\cite{25ZhangWhich} achieve below 53\% accuracy on fault type and below 56\% on fault step (Table~\ref{tab:fault_diagnosis}).
Beyond these results, our evaluation reveals several new findings (discussed in \S\ref{sec:discussion}).
(1)~\textit{The most severe faults are not the most harmful.}
(2)~\textit{The most harmful faults are also the hardest to diagnose.}
(3)~\textit{Robustness depends on system implementation.}

In summary, we make the following contributions.

\begin{itemize}[leftmargin=*, nosep]
\item We define a fault taxonomy for LLM API responses by adapting the classical fault classification from distributed systems~\cite{04AvizienisBasic}. The taxonomy covers crash, omission, and value faults on both content and tool call fields. Combined with injection strategies and compound scenarios, it yields 65 fault configurations (\S\ref{sec:fault_classification}, \S\ref{sec:method_config}).

\item We design a controlled, runtime, non-intrusive fault injection framework for agent systems. The framework installs a wrapper at the HTTP layer to intercept and modify LLM API responses at runtime without modifying any agent system source code (\S\ref{sec:method_injection}), and verifies whether each fault was actually triggered to filter untriggered tasks from evaluation (\S\ref{sec:method_verification}).

\item We evaluate across multiple agent systems, benchmarks, backbone LLMs, and fault configurations. All systems degrade under fault injection (pass@1 drops up to 50 percentage points), and existing fault diagnosis methods achieve below 56\% accuracy.
\end{itemize}


\section{Background}\label{sec:Background}

\subsection{Agent Systems and LLM API Dependence}\label{sec:bg_mas}

An agent system uses one or more LLM-powered agents to solve complex tasks~\cite{25YehudaiAgents, 24WuAutoGen, 25ChenSecureAgentBench, 26WangNativeBench, 26WangOpsBench}. Each agent has a role (e.g., coder, reviewer, planner) and calls the LLM API to generate text, invoke tools (e.g., code executors, file editors, web search), or communicate with other agents through message passing. Each activated agent makes at least one LLM API call, so a task involves a chain of LLM API calls and tool invocations across agents and rounds.

Agent systems access LLMs through HTTP-based APIs. Providers expose endpoints such as OpenAI \texttt{/v1/chat/completions} and Anthropic \texttt{/v1/messages}, but all follow the same HTTP request-response pattern. An agent sends conversation history and receives the LLM output. This shared communication layer enables our non-intrusive fault injection. Although providers use JSON schemas, their responses contain generated text, tool call arguments, and completion metadata. Most agent frameworks use OpenAI-compatible APIs, so we use the OpenAI Chat Completions format throughout this paper. As shown in Figure~\ref{fig:response_example}, its \texttt{message} object contains two mutually exclusive fields. The \texttt{content} field stores generated text, while \texttt{tool\_calls} stores structured tool arguments. The \texttt{finish\_reason} field indicates normal completion (\texttt{stop}), token-limit truncation (\texttt{length}), or tool invocation (\texttt{tool\_calls}).


\subsection{Fault Injection and Chaos Engineering}

Fault injection is a technique that introduces controlled faults into a system to evaluate its robustness. Chaos engineering systematizes this idea into a discipline that proactively injects faults into running systems to discover weaknesses before they cause failures in production~\cite{26YuFailure, 25MaAIOpsLab, 24ChenMicroFI, 25HuangConan}. The core principle is that a system's fault tolerance cannot be verified under normal conditions alone. Only by injecting realistic faults into a running system can developers observe how the system responds, such as retries, path switches, or silent error propagation, and identify weaknesses that would otherwise remain hidden. In traditional software systems, tools such as ChaosBlade~\cite{26ChaosBlade} inject infrastructure-level faults (e.g., instance termination, CPU exhaustion, network delay) to test robustness.

\section{Fault Taxonomy for LLM API Responses}\label{sec:fault_classification}

\begin{figure}[t]
\centering
\begin{lstlisting}[style=codestyle, basicstyle=\ttfamily\scriptsize, escapechar=|]
|\textcolor{gray}{// (a) Text generation}|
{"choices": [{"message": {
    |\colorbox{blue!6}{\textcolor{blue!70!black}{"content": "def add(a, b):$\backslash$n~~~~return a + b"}}|,
    |\textcolor{gray!70}{"tool\_calls": null}|},
  "finish_reason": |\textcolor{teal!70!black}{"stop"}|}]}

|\textcolor{gray}{// (b) Tool invocation}|
{"choices": [{"message": {
    |\textcolor{gray!70}{"content": null}|,
    |\colorbox{orange!6}{\textcolor{orange!60!black}{"tool\_calls": [\{"function": \{"name": "execute",}}|
    |\colorbox{orange!6}{\textcolor{orange!60!black}{~"arguments": "\{$\backslash$"code$\backslash$":$\backslash$"print(add(1,2))$\backslash$"\}"\}\}]}}|},
  "finish_reason": |\textcolor{teal!70!black}{"tool\_calls"}|}]}
\end{lstlisting}
\vspace{-7pt}
\caption{OpenAI Chat Completions API response. \textcolor{blue!70!black}{\texttt{content}} and \textcolor{orange!60!black}{\texttt{tool\_calls}} are mutually exclusive. Inactive fields are \textcolor{gray!70}{gray}. \textcolor{teal!70!black}{\texttt{finish\_reason}}: finish metadata.}
\label{fig:response_example}
\vspace{-14pt}
\end{figure}

\begin{table}[t]
\centering
\caption{Fault taxonomy for LLM API responses. \ding{51} marks a valid target field.}
\label{tab:fault_taxonomy}
\vspace{-7pt}
{\setlength{\tabcolsep}{3pt}
\resizebox{\linewidth}{!}{
\begin{tabular}{llccl}
\Xhline{1pt}
\textbf{Category} & \textbf{Fault type} & \textbf{Content} & \textbf{Tool call} & \textbf{Real-world scenario} \\
\Xhline{1pt}
\multirow{2}{*}{Crash} & Error & \ding{51} & \ding{51} & server overload, HTTP 5xx, rate limiting~\cite{26ZhangEngineering,26AgentFrameworkBugs} \\
& Timeout & \ding{51} & \ding{51} & network congestion, backend delay, API latency~\cite{26ZhangEngineering,26ShaoComfrey} \\
\hline
\multirow{2}{*}{Omission} & Empty & \ding{51} & \ding{51} & safety filter, content policy rejection~\cite{26AzureContentFilter} \\
& Truncate & \ding{51} & \ding{51} & token limit, TCP interruption, incomplete completion~\cite{26ZhangEngineering,26AgentFrameworkBugs} \\
\hline
\multirow{2}{*}{Value} & Corrupt & \ding{51} & \ding{51} & encoding error, garbled characters~\cite{26YuWhyDoes} \\
& Schema & \ding{51} & \ding{51} & parsing error, schema mismatch~\cite{25SongCallNavi,26SigdelSchemaFirst,26AgentFrameworkBugs} \\
\Xhline{1pt}
\end{tabular}
}}
\vspace{-14pt}
\end{table}

To address the lack of a systematic fault taxonomy for LLM API responses, we make a deep investigation on fault types at the LLM API response layer. Using a deductive classification based on dependability theory~~\cite{04AvizienisBasic}, several authors take the three fault categories (i.e., crash, omission, value), apply each to the response fields in \S\ref{sec:bg_mas}, and then discuss and split each into finer types: a crash fault yields Error and Timeout, an omission fault yields Empty and Truncate, a value fault yields Corrupt and Schema. Since an LLM API response has a fixed set of fields, this enumeration covers all fault types rather than selecting them by hand. Existing fault taxonomies for agent systems and ours both describe the faults that can affect agent systems, but differ in sources and layers. They group observed faults through developer surveys~\cite{niu2026trust}, incident analysis~\cite{yang2023users}, or literature review~\cite{26JiaMASFIRE} at the prompt or application layer, while ours is fixed and targets the response layer. Thus, these two taxonomies are complementary. An LLM API response has a fixed structure with a known set of fields (\S\ref{sec:bg_mas}). Therefore, this investigation produces a complete fault list that covers every way a response can deviate for each category without relying on human experience, and the same process applies when new API fields appear in the future.
Faults in our taxonomy are not hypothetical. Empirical studies on LLM-based software in production report that API errors, connection timeouts, rate limiting, and response format issues are among the most common failures in agent frameworks~\cite{26AgentFrameworkBugs}, AI coding tools~\cite{26ZhangEngineering}, and LLM-integrated applications~\cite{26ShaoComfrey,25ShaoHydrangea}, and our taxonomy systematically covers all of them.

\noindent
\textbf{Target fields.}
The first dimension of our investigation is the target field. As described in \S\ref{sec:bg_mas}, \texttt{content} holds generated text and \texttt{tool\_calls} holds structured tool arguments. Faults on \texttt{content} corrupt the text that the next agent reads. Faults on \texttt{tool\_calls} corrupt the arguments passed to tools. Because \texttt{content} is a plain text string while \texttt{tool\_calls} contains structured JSON, the same fault type affects each field differently. We apply each fault type to both fields to produce the configurations in Table~\ref{tab:fault_taxonomy}. This distinction matters in practice because tool invocation and parameter generation are known sources of failures in agent systems that rely on LLM APIs~\cite{25SongCallNavi,26SigdelSchemaFirst,26AgentFrameworkBugs}.

\noindent
\textbf{Fault categories.}
The second dimension is the fault category. In distributed systems~\cite{04AvizienisBasic}, faults are classified by the severity of service deviation. A \textit{crash fault} returns no valid response, an \textit{omission fault} returns an incomplete response, and a \textit{value fault} returns a response with incorrect content. For traditional software services, faults on structured outputs are easy to detect by checking error codes or field values. LLM API responses are different because they contain free-form text such as reasoning, code, or tool call arguments, and the expected output is not predefined. A fault on such content, such as truncated code or garbled characters, cannot be detected by checking any single field. This makes LLM API faults harder to detect and more likely to propagate silently. A faulty LLM response still looks normal, so the receiving agent uses it as normal input and passes it to the next agent. This differs from a traditional distributed system, where a fault surfaces through an error code and is caught where it occurs. We adapt each fault category to the LLM API response structure as described below.

\noindent
\textbf{Fault types.}
We define six fault types across the three categories. Table~\ref{tab:fault_taxonomy} maps each type to its real deployment scenario: Error to server overload and HTTP 5xx, Timeout to network delay, Empty to a safety filter, Truncate to a token-limit cutoff, Corrupt to an encoding error, and Schema to a tool-call format mismatch.

\textbf{\textit{Crash faults}} make the response unusable for the receiving agent.
\begin{itemize}[leftmargin=*, nosep]
\item \textbf{Error.} The LLM API returns an error message instead of the expected output, as happens during server overload, deployment rollouts, or rate limiting~\cite{26ZhangEngineering,26AgentFrameworkBugs}.
\item \textbf{Timeout.} The LLM API does not respond within the allowed time, as happens during network congestion or backend processing delays~\cite{26ZhangEngineering,26ShaoComfrey}.
\end{itemize}

\textbf{\textit{Omission faults}} return a valid response structure with missing content.
\begin{itemize}[leftmargin=*, nosep]
\item \textbf{Empty.} The LLM API returns a response with no content, as happens when a safety filter or content policy blocks the output~\cite{26AzureContentFilter}. The response parses but contains no useful information.
\item \textbf{Truncate.} The LLM API returns a response that is cut off partway through, as happens when the token limit is reached or a TCP connection drops during streaming~\cite{26ZhangEngineering,26AgentFrameworkBugs}. The agent receives a partial response that may look like a short but complete answer, making truncation harder to detect than an empty response.
\end{itemize}

\textbf{\textit{Value faults}} return a response that is structurally valid and complete, but with wrong content. These are the hardest faults to detect because the response looks normal from the outside. Semantic faults such as hallucination and semantic drift also fall in this category, since the model returns a complete response whose content is wrong. They originate in the model rather than in the API layer, so our framework classifies them here but does not inject them.
\begin{itemize}[leftmargin=*, nosep]
\item \textbf{Corrupt.} The response content is damaged by encoding errors, as happens when proxy servers or caching layers apply incorrect character encoding. The output parses but is semantically meaningless.
\item \textbf{Schema.} The response contains valid JSON that does not follow the expected structure, as happens when the LLM generates malformed output or a parsing error corrupts the response~\cite{25SongCallNavi,26SigdelSchemaFirst,26AgentFrameworkBugs}. No parsing error is raised, and the fault propagates silently.
\end{itemize}

This taxonomy defines what faults to inject when we evaluate how robust agent systems are under LLM API faults. Combined with injection strategies and compound scenarios (\S\ref{sec:method_config}), it expands into the fault configurations used in the experiments, letting us measure the degradation each fault causes and identify which fault type and configuration degrade each architecture most.
\begin{figure*}[t]
\centering
\includegraphics[width=0.79\linewidth]{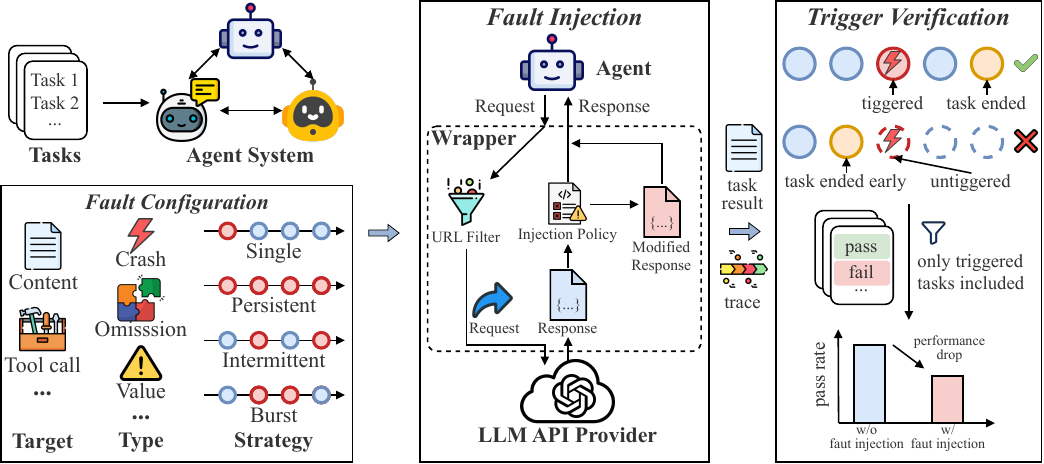}
\vspace{-7pt}
\caption{Overview of \ourmethod. The framework takes an agent system, tasks, and a fault configuration as input, injects faults at the HTTP layer, and outputs task results with execution traces for robustness evaluation.}
\label{fig:overview}
\vspace{-7pt} 
\end{figure*}

\section{Methodology}\label{sec:Methodology}

\subsection{Overview}

We present \ourmethod, a chaos engineering framework that evaluates the robustness of agent systems through controlled, runtime, non-intrusive fault injection at the LLM API layer. All agent systems communicate with LLM services through HTTP requests, regardless of their internal architecture (\S\ref{sec:bg_mas}). We exploit this shared layer by installing a fault injection wrapper on the HTTP client at runtime. The wrapper intercepts every LLM API response, decides whether to inject a fault based on the configured policy, modifies the response if the policy fires, and returns the result to the agent system. No source code of the agent system is changed.

As illustrated in Fig.~\ref{fig:overview}, the framework takes an agent system, tasks, and a fault configuration as input. It outputs task results with execution traces that record every LLM API call. The framework has the following components: (1)~\textit{fault configuration} (\S\ref{sec:method_config}) combines fault types, target fields, injection strategies, and compound scenarios into complete configurations, (2)~\textit{fault injection} (\S\ref{sec:method_injection}) implements the HTTP layer wrapper that intercepts and modifies LLM API responses at runtime, and (3)~\textit{trigger verification} (\S\ref{sec:method_verification}) checks execution traces after task completion and filters tasks where the configured fault was not triggered.

\vspace{-10pt}
\subsection{Fault Configuration}\label{sec:method_config}

The fault taxonomy (\S\ref{sec:fault_classification}) defines the fault types and target fields. To form complete configurations, we further specify injection frequency, compound combinations, and injection position.

\textbf{Injection strategies.} Different real failures affect different numbers of LLM calls within a single task. A network glitch may corrupt only one call, while an expired API key causes every call to fail. We define the following strategies to cover these patterns.

\begin{itemize}[leftmargin=*, nosep]
\item \textit{Single}. Inject the fault once at the first matching LLM call, then stop. This models a transient failure such as a network glitch.
\item \textit{Persistent}. Inject the fault at every matching LLM call throughout the entire task. This models a sustained failure such as an expired API key or a misconfigured proxy.
\item \textit{Intermittent}. Inject the fault at each matching LLM call independently with probability 0.3. This models a flaky connection where roughly 30\% of requests fail.
\item \textit{Burst}. Inject the fault at the first 3 consecutive matching LLM calls, then stop. This models a temporary overload where the server rejects requests briefly and then recovers.
\end{itemize}

\textbf{Compound scenarios.} Real deployment failures can change the response in one or more ways. For example, server overload may first add a delay and then return an error response. A content safety filter may strip tool calls and replace the content with a rejection message simultaneously. To model such cases, we define compound scenarios that reproduce a real deployment failure by modifying the same LLM API response within a single interception. A scenario is grouped by the failure it reproduces, not by the number of fault types, so it stays compound even when it maps to a single type. We take these failures from empirical studies on LLM-based software~\cite{26ShaoComfrey,25ShaoHydrangea,26ZhangEngineering,26AgentFrameworkBugs} and vendor documentation~\cite{26AzureContentFilter}, selecting those we can reproduce at the LLM API response layer. The fault types column maps each failure to the base types from our taxonomy (\S\ref{sec:fault_classification}) that its response changes most resemble. Table~\ref{tab:fault_compound} lists all compound scenarios. Each scenario applies the changes its failure produces, which may affect one or several fields of the response.

\textbf{Configuration space.} A complete fault configuration consists of four elements: (1)~the fault type (from \S\ref{sec:fault_classification}), (2)~the target field (\texttt{content} or \texttt{tool\_calls}), (3)~the injection strategy (single, persistent, intermittent, or burst), and (4)~the injection position (which LLM call to start injecting at). We construct the full set of configurations in three groups. \textit{Basic experiments} combine each of the 6 fault types with each of the 2 target fields and each of the 4 injection strategies, yielding $6 \times 2 \times 4 = 48$ configurations. \textit{Position experiments} inject selected fault types (error, timeout, schema) at different call positions (1st, 2nd, 3rd call), yielding $3 \times 3 = 9$ configurations. \textit{Compound experiments} use the 8 scenarios in Table~\ref{tab:fault_compound}. Adding up the three groups gives $48 + 9 + 8 = 65$ fault configurations.


\begin{table}[t]
\centering
\caption{Compound fault scenarios. Each reproduces one real deployment failure.}
\label{tab:fault_compound}
\vspace{-7pt}
\resizebox{\linewidth}{!}{
\begin{tabular}{lll}
\Xhline{1pt}
\textbf{Scenario} & \textbf{Fault Types} & \textbf{Description} \\
\Xhline{1pt}
API degradation & Timeout, Error & Delay then return error response \\
Content filter & Empty, Schema & Remove tool calls and replace content with filter message \\
Max tokens & Truncate & Truncate content and set finish reason to length \\
Proxy HTML & Corrupt & Replace content with an HTML error page \\
Stale cache & Corrupt & Replay previous response on the next call \\
Stale data & Schema & Replace tool call arguments with wrong values \\
Wrong entity & Schema & Replace tool call arguments with ambiguous values \\
Slow response & Timeout & Add delay with no content change \\
\Xhline{1pt}
\end{tabular}
}
\vspace{-7pt}
\end{table}

\subsection{Non-Intrusive Fault Injection}\label{sec:method_injection}

\begin{figure}[t]
\centering
\includegraphics[width=\linewidth]{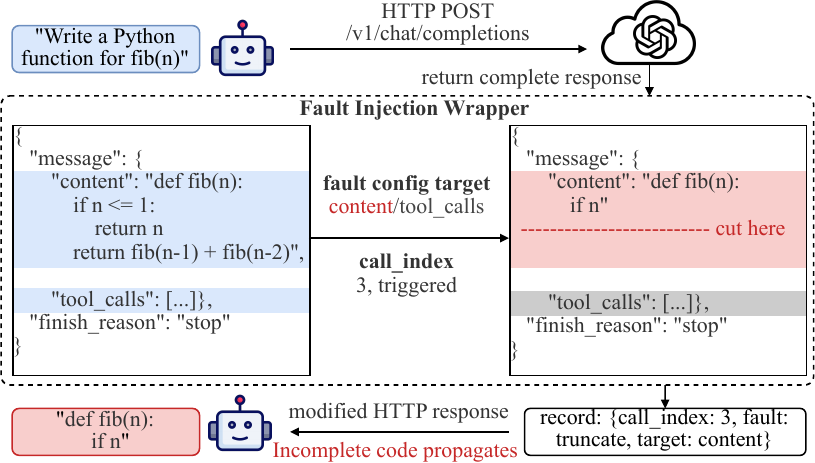}
\vspace{-14pt}
\caption{Fault injection wrapper. It intercepts LLM API responses, injects the configured fault, records the change in the execution trace, and returns the response to the agent.}
\label{fig:wrapper}
\vspace{-14pt} 
\end{figure}

To address the offline and intrusive limitations of existing injection methods, we install a fault injection wrapper at the HTTP layer shared by all agent systems (\S\ref{sec:bg_mas}). The wrapper provides controlled fault injection at runtime and requires no modification to any agent system source code.

\textbf{Wrapper installation.} Before execution starts, we replace the HTTP client's request method with the wrapper, so that every outgoing request passes through the wrapper before reaching the LLM provider. To support concurrent execution in which multiple tasks run in parallel with different fault configurations, each task is bound to a separate injection state, ensuring that fault sequences are isolated across tasks. In our implementation, we use monkey patching on \texttt{httpx.AsyncClient} and Python's \texttt{contextvars} for per-task state isolation.

\textbf{Request interception.} At each HTTP request, the wrapper performs the following steps (Fig.~\ref{fig:wrapper}). First, it checks whether the request URL matches an LLM API endpoint (e.g., \texttt{/chat/completions}). Requests to other URLs pass through unchanged. Second, the wrapper ensures that the full response body is available before modification. Third, the wrapper forwards the request to the real LLM provider and receives the complete response. Fourth, it evaluates the injection policy (described below) to decide whether to inject a fault at this call, and applies the modification if the policy fires. Finally, the wrapper records the call in the execution trace and returns the (possibly modified) response to the agent system.

\textbf{Injection policy.} The injection policy decides whether to inject a fault at a given LLM call. The wrapper maintains a call counter and a fire counter per task. The call counter increments each time the wrapper intercepts an LLM API request. The fire counter tracks how many times the fault has actually been injected so far.

Before checking the injection strategy, the wrapper applies a target field guard. Since \texttt{content} and \texttt{tool\_calls} are mutually exclusive (\S\ref{sec:bg_mas}), the wrapper checks whether the response contains the configured target field. If not, it skips this call and lets the response pass through without incrementing the fire counter. This guard ensures that faults are applied only to responses that contain the intended target field.

If the target field matches, the wrapper evaluates the injection strategy using the call counter and fire counter. For \textit{single}, the policy fires once (when the fire counter is 0 and the call counter has reached the configured start position), then deactivates. For \textit{persistent}, the policy fires on every matching call from the start position onward. For \textit{intermittent}, the policy fires on each matching call with the configured probability (default 0.3), which is evaluated independently per call for reproducibility. For \textit{burst}, the policy fires on consecutive matching calls while the fire counter is below the configured burst count (default 3), then deactivates.

\textbf{Modification functions.} Each fault type maps to a deterministic modification function that operates on the response JSON at \texttt{choices[0].message}. We describe how each function modifies the \texttt{content} field (plain text) and the \texttt{tool\_calls} field (structured JSON parsed, modified recursively, and serialized back) respectively.

For \textit{crash faults}, \textbf{error} replaces the \texttt{content} field with an error like ``HTTP 500 Internal Server Error'', and replaces \texttt{tool\_calls} arguments with an empty object. \textbf{Timeout} replaces the \texttt{content} field with a timeout notification and clears \texttt{tool\_calls}.

For \textit{omission faults}, \textbf{empty} sets the \texttt{content} field to an empty string, or the \texttt{tool\_calls} field to an empty list. \textbf{Truncate} keeps the first 30\% of the target field (configurable). For \texttt{content}, it also sets \texttt{finish\_reason} to \texttt{``length''} to match a token limit cutoff. For \texttt{tool\_calls}, it parses the JSON arguments, truncates all string values to 30\% of their length, and serializes the result to valid JSON.

For \textit{value faults}, \textbf{corrupt} applies UTF-8 to Latin-1 misinterpretation on the \texttt{content} field, producing visually garbled but structurally intact text. For \texttt{tool\_calls}, it replaces approximately 20\% of characters in all string values with random Unicode symbols. \textbf{Schema} replaces the \texttt{content} field with a JSON error object (e.g., \texttt{\{``error'', ``content\_policy\_violation''\}}), and replaces \texttt{tool\_calls} arguments with a JSON object containing unexpected keys (e.g., \texttt{\{``wrong\_param'', ``unexpected\_value''\}}). In real deployments an LLM can return valid JSON whose keys do not match the tool's expected schema, a failure that tool-call studies report as common~\cite{25SongCallNavi,26SigdelSchemaFirst,26AgentFrameworkBugs}. We reproduce this failure with a fixed placeholder key, since the mismatch is what matters, not the key name.

All modification functions are deterministic given the same configuration and response, and each keeps the response body parseable as valid JSON. The wrapper also caches each original response to support the replay action used in compound scenarios such as stale cache (Table~\ref{tab:fault_compound}).

\textbf{Trace recording.} After each interception, the wrapper records the call as a span in the execution trace, covering every LLM API call in the task rather than only the injected one. Each span contains the call index, the request content, the original response, and the modified response (if a fault was applied). When the wrapper applies a fault, it additionally logs a fault event that records the call index, fault type, and target field. This trace provides the complete record needed for trigger verification (\S\ref{sec:method_verification}) and fault diagnosis (\S\ref{sec:rq3}).

\vspace{-7pt}
\subsection{Trigger Verification}\label{sec:method_verification}

An agent system decides how many LLM calls to make at runtime, and this number varies across tasks. A fault configured for a specific call position may not fire if the task finishes before reaching that position. For example, if a fault is configured for the 3rd LLM call but the task finishes in 2 calls, the fault is never applied. The intermittent strategy may also not select any call during a short task. The target field guard (\S\ref{sec:method_injection}) may cause additional skips if the LLM consistently returns tool calls when a content fault is configured. Including such untriggered tasks in the evaluation would mix faulted and unfaulted results, making the system appear more robust than it actually is.

We therefore check after each task whether the execution trace contains at least one fault event (as recorded by the wrapper in \S\ref{sec:method_injection}). If no fault event exists, the task is marked as untriggered and excluded from the evaluation. Only triggered tasks are used to compute $\text{pass@1}_{\text{w/ FI}}$ and $\Delta$pass@1. This filtering ensures that the measured degradation reflects the true impact of the injected fault.

\section{Experimental Evaluation}

In this section, we evaluate \ourmethod by answering the following research questions:
\begin{itemize}[leftmargin=*, nosep]
\item \textbf{RQ1:} How robust are agent systems under LLM API faults?
\item \textbf{RQ2:} How do different fault configurations affect robustness?
\item \textbf{RQ3:} How effectively can existing methods diagnose the injected fault type and step?
\end{itemize}

\subsection{Experiment Setup}\label{sec:ExperimentSetup}

\textbf{Agent Systems.}
We evaluate agent systems covering diverse architectural patterns. We select one public system for each structurally distinct pattern: AutoGen for conversation, MAD for debate, MapCoder for pipeline, EvoMAC for evolutionary, and Mini-SE for single-agent. Production agents build on the same patterns~\cite{25OpenAIChatGPTAgent,25AnthropicClaudeCode}, so these systems cover the main agent styles. To ensure fair comparison and uniform fault injection, we reimplement all systems on Google ADK~\cite{26GoogleADK} with unified tool interfaces, preserving each system's original interaction logic while standardizing the underlying LLM API layer.
\begin{itemize}[leftmargin=*, nosep]
\item \textbf{AutoGen}~\cite{24WuAutoGen} uses a conversation pattern in which an assistant and a user proxy iteratively generate and execute code.
\item \textbf{MAD}~\cite{24LiangMAD} uses a debate pattern in which agents argue across rounds under a moderator before a judge selects the final answer.
\item \textbf{MapCoder}~\cite{24IslamMapCoder} uses a pipeline pattern in which a retriever, planner, coder, verifier, and debugger run in fixed stages.
\item \textbf{EvoMAC}~\cite{25HuEvoMAC} uses an evolutionary pattern in which agents decompose tasks and refine code across generations.
\item \textbf{Mini-SE}~\cite{25XuScalable} uses an agent to search, view, and edit files and submit patches for multi-file software issues.
\end{itemize}


\begin{table*}[t]
\centering
\caption{Pass@1 without fault injection (w/o) and with fault injection (w/), and $\Delta$pass@1 ($\Delta$) for each model, system, and dataset. Mini-SE is evaluated only on SWE-bench Pro. \textbf{Bold}: highest $\Delta$pass@1 per dataset within each model.}
\label{tab:overall}
\vspace{-7pt}
\renewcommand{\arraystretch}{1.05}
\setlength{\tabcolsep}{2.0pt}
\resizebox{\textwidth}{!}{
\begin{tabular}{ll|ccc|ccc|ccc|ccc|ccc|ccc|ccc}
\Xhline{1.2pt}
\rowcolor{gray!12}
& & \multicolumn{3}{c|}{\textbf{HumanEval}} & \multicolumn{3}{c|}{\textbf{HumanEval+}} & \multicolumn{3}{c|}{\textbf{MBPP}} & \multicolumn{3}{c|}{\textbf{MBPP+}} & \multicolumn{3}{c|}{\textbf{MMLU-Pro}} & \multicolumn{3}{c|}{\textbf{MATH-500}} & \multicolumn{3}{c}{\textbf{SWE-bench Pro}} \\
\rowcolor{gray!12}
\multirow{-2}{*}{\textbf{Model}} & \multirow{-2}{*}{\textbf{System}}
& \textbf{w/o} & \textbf{w/} & $\boldsymbol{\Delta}$
& \textbf{w/o} & \textbf{w/} & $\boldsymbol{\Delta}$
& \textbf{w/o} & \textbf{w/} & $\boldsymbol{\Delta}$
& \textbf{w/o} & \textbf{w/} & $\boldsymbol{\Delta}$
& \textbf{w/o} & \textbf{w/} & $\boldsymbol{\Delta}$
& \textbf{w/o} & \textbf{w/} & $\boldsymbol{\Delta}$
& \textbf{w/o} & \textbf{w/} & $\boldsymbol{\Delta}$ \\
\Xhline{1.2pt}
& AutoGen  & 98.61 & 79.17 & 19.44 & 98.59 & 77.46 & 21.13 & 91.35 & 74.04 & 17.31 & 74.19 & 62.58 & 11.61 & 58.97 & 51.92 & 7.05 & 29.32 & 20.94 & 8.38 & -- & -- & -- \\
& MAD      & 83.44 & 59.24 & 24.2 & 82.8 & 57.96 & 24.84 & 83.67 & 59.18 & 24.49 & 58.66 & 43.58 & 15.08 & 65.84 & 45.2 & 20.64 & 68.42 & 47.72 & 20.7 & -- & -- & -- \\
& MapCoder & 96.53 & 47.92 & \textbf{48.61} & 97.22 & 47.92 & \textbf{49.3} & 99.55 & 58.48 & \textbf{41.07} & 75.61 & 34.76 & \textbf{40.85} & 76.96 & 38.71 & \textbf{38.25} & 60.48 & 26.21 & \textbf{34.27} & -- & -- & -- \\
& EvoMAC   & 98.73 & 80.25 & 18.48 & 98.7 & 80.52 & 18.18 & 97.5 & 80.83 & 16.67 & 78.19 & 63.46 & 14.73 & 75.17 & 61.54 & 13.63 & 64.79 & 48.94 & 15.85 & -- & -- & -- \\
\multirow{-5}{*}{\llmname{Claude-Sonnet-4.5}}
& Mini-SE  & -- & -- & -- & -- & -- & -- & -- & -- & -- & -- & -- & -- & -- & -- & -- & -- & -- & -- & 11.3 & 10.43 & \textbf{0.87} \\
\hline
& AutoGen  & 95.71 & 74.29 & 21.42 & 100 & 77.14 & 22.86 & 92.23 & 73.79 & 18.44 & 77.48 & 60.93 & 16.55 & 48.78 & 37.4 & 11.38 & 68.03 & 55.74 & 12.29 & -- & -- & -- \\
& MAD      & 87.83 & 69.57 & 18.26 & 90.16 & 68.85 & 21.31 & 86.6 & 68.04 & 18.56 & 63.06 & 54.48 & 8.58 & 41.36 & 30.37 & 10.99 & 50.93 & 40.28 & 10.65 & -- & -- & -- \\
& MapCoder & 91.89 & 42.57 & \textbf{49.32} & 93.96 & 44.3 & \textbf{49.66} & 99.57 & 56.65 & \textbf{42.92} & 69.76 & 26.95 & \textbf{42.81} & 33.88 & 14.69 & \textbf{19.19} & 31.56 & 11.03 & \textbf{20.53} & -- & -- & -- \\
& EvoMAC   & 89.22 & 65.69 & 23.53 & 87.38 & 61.17 & 26.21 & 84.42 & 64.94 & 19.48 & 67.26 & 50.67 & 16.59 & 52.33 & 43.01 & 9.32 & 48.45 & 40.72 & 7.73 & -- & -- & -- \\
\multirow{-5}{*}{\llmname{GPT-5.2}}
& Mini-SE  & -- & -- & -- & -- & -- & -- & -- & -- & -- & -- & -- & -- & -- & -- & -- & -- & -- & -- & 3.47 & 6.18 & \textbf{$-$2.71} \\
\hline
& AutoGen  & 89.42 & 74.04 & 15.38 & 91.26 & 77.67 & 13.59 & 89.22 & 74.25 & 14.97 & 66.06 & 57.3 & 8.76 & 55.8 & 58.04 & $-$2.24 & 57.89 & 39.04 & 18.85 & -- & -- & -- \\
& MAD      & 41.13 & 23.4 & 17.73 & 25.17 & 14.97 & 10.2 & 24.77 & 15.77 & 9 & 40.06 & 28.39 & 11.67 & 29.39 & 18.7 & 10.69 & 30.35 & 20.62 & 9.73 & -- & -- & -- \\
& MapCoder & 92.96 & 46.48 & \textbf{46.48} & 93.71 & 45.45 & \textbf{48.26} & 100 & 59.38 & \textbf{40.62} & 71.21 & 28.79 & \textbf{42.42} & 50.37 & 24.81 & \textbf{25.56} & 41.83 & 17.87 & \textbf{23.96} & -- & -- & -- \\
& EvoMAC   & 96.69 & 76.82 & 19.87 & 97.42 & 78.71 & 18.71 & 95.34 & 78.81 & 16.53 & 68.56 & 53.89 & 14.67 & 54.22 & 42.57 & 11.65 & 47.62 & 38.1 & 9.52 & -- & -- & -- \\
\multirow{-5}{*}{\llmname{DeepSeek-V3.2}}
& Mini-SE  & -- & -- & -- & -- & -- & -- & -- & -- & -- & -- & -- & -- & -- & -- & -- & -- & -- & -- & 5.96 & 6.88 & \textbf{$-$0.92} \\
\hline
& AutoGen  & 94.67 & 74.67 & 20.0 & 98.63 & 73.97 & 24.66 & 93.4 & 76.42 & 16.98 & 71.24 & 58.17 & 13.07 & 35.61 & 29.55 & 6.06 & 24.46 & 24.46 & 0 & -- & -- & -- \\
& MAD      & 75.16 & 56.05 & 19.11 & 75.0 & 57.69 & 17.31 & 73.58 & 59.76 & 13.82 & 69.41 & 60.06 & 9.35 & 29.93 & 27.82 & 2.11 & 50.35 & 39.79 & 10.56 & -- & -- & -- \\
& MapCoder & 89.93 & 43.17 & \textbf{46.76} & 92.09 & 46.04 & \textbf{46.05} & 99.08 & 58.06 & \textbf{41.02} & 66.47 & 29 & \textbf{37.47} & 38.22 & 16.99 & \textbf{21.23} & 30.65 & 14.18 & \textbf{16.47} & -- & -- & -- \\
& EvoMAC   & 96.58 & 76.03 & 20.55 & 95.92 & 77.55 & 18.37 & 94.82 & 75.65 & 19.17 & 70.79 & 53.26 & 17.53 & 47.24 & 36.21 & 11.03 & 45.49 & 35.02 & 10.47 & -- & -- & -- \\
\multirow{-5}{*}{\llmname{Seed-1.8}}
& Mini-SE  & -- & -- & -- & -- & -- & -- & -- & -- & -- & -- & -- & -- & -- & -- & -- & -- & -- & -- & 5.7 & 4.39 & \textbf{1.31} \\
\Xhline{1.2pt}
\end{tabular}
}
\vspace{-7pt}
\end{table*}

\noindent
\textbf{Datasets.}
We use benchmarks for code generation, knowledge reasoning, mathematical reasoning, and repository-level bug fixing. AutoGen, MAD, MapCoder, and EvoMAC are evaluated on all benchmarks except SWE-bench Pro, which requires specialized software engineering tools unavailable to these systems. Mini-SE is evaluated only on SWE-bench Pro. To reduce computational cost, we randomly sample 300 examples from large datasets and use all examples from smaller datasets.
\begin{itemize}[leftmargin=*, nosep]
\item \textbf{HumanEval}~\cite{21ChenHumanEval} contains manually crafted Python problems with test cases for correctness.
\item \textbf{MBPP}~\cite{21AustinMBPP} contains crowd-sourced Python problems, each with a requirement, function signature, and test cases.
\item \textbf{HumanEval+ / MBPP+}~\cite{23LiuHumanEvalPlus} are EvalPlus extensions of HumanEval and MBPP with more tests for stricter evaluation.
\item \textbf{MMLU-Pro}~\cite{24WangMMLUPro} contains curated reasoning questions from textbooks and exams across domains.
\item \textbf{MATH-500}~\cite{24WangMATH500} contains challenging math problems that require multi-step reasoning.
\item \textbf{SWE-bench Pro}~\cite{25DengSWEBenchPro} contains long-horizon software engineering tasks that often require patches across files.
\end{itemize}

\noindent
\textbf{Fault assignment.}
We construct 65 fault configurations from the taxonomy. (1)~\textit{Basic experiments} combine fault types, target fields, and injection strategies. (2)~\textit{Position experiments} inject faults at different call positions. (3)~\textit{Compound experiments} combine multiple fault types in a single execution. We uniformly distribute configurations across tasks, so each task receives one configuration.

\noindent
\textbf{Metrics.}
We measure task success using pass@1~\cite{21ChenHumanEval}, the percentage of tasks solved correctly with one attempt.
We run each system under two conditions, without fault injection (w/o FI) and with fault injection (w/ FI).
We quantify robustness degradation by $\Delta$pass@1, the difference in pass@1 before and after fault injection.
A higher $\Delta$pass@1 indicates lower robustness, where $\Delta\text{pass@1} = \text{pass@1}_{\text{w/o FI}} - \text{pass@1}_{\text{w/ FI}}$.

We compute $\text{pass@1}_{\text{w/ FI}}$ only over tasks where the fault was actually triggered, as described in \S\ref{sec:method_verification}.
For robustness evaluation, we report pass@1 and $\Delta$pass@1.
For fault diagnosis, we report \textbf{type accuracy} and \textbf{step accuracy}, where the ground truth is the injected fault type and the step index of the first injection. Type accuracy is the percentage of failed tasks where the predicted type matches the ground truth. Step accuracy is the percentage where the predicted step matches.

\noindent
\textbf{Implementation.}
We implement \ourmethod in Python 3.12 and test with LLMs from different providers: \llmname{Claude-Sonnet-4.5} \cite{25Claude45}, \llmname{GPT-5.2}~\cite{25CHATGPT}, \llmname{DeepSeek-V3.2}~\cite{25DeepSeekV32}, and \llmname{Seed-1.8}~\cite{25Seed}, all with temperature 0.7, because it is a common default in existing studies~\cite{25WangMoA,25ChenSWEExp}. Code execution tools run in sandboxed subprocesses with a 30-second timeout. Each task allows up to 50 LLM calls for multi-agent systems and 100 for Mini-SE. Experiments run on Ubuntu 24.04 with Intel Xeon Gold 6326 CPU and 128GB RAM.

\subsection{RQ1: Robustness of Agent Systems}\label{sec:rq1}
Table~\ref{tab:overall} presents pass@1 and $\Delta$pass@1 for all models, systems, and datasets, while Table~\ref{tab:overhead} reports resource consumption. Unless otherwise noted, analyses use \llmname{Claude-\allowbreak{Sonnet-4.5}}.

\textbf{Overall degradation.} Table~\ref{tab:overall} shows that fault injection lowers pass@1 for all systems on most datasets. $\Delta$pass@1 ranges from 0.87\% (Mini-SE on SWE-bench Pro) to 49.66\% (MapCoder on HumanEval+ under \llmname{GPT-5.2}). Even AutoGen, with the lowest $\Delta$pass@1 among these systems, drops from 98.61\% to 79.17\% on HumanEval ($\Delta$=19.44\%). These results confirm that LLM API faults degrade performance across systems and datasets. These numbers are measured only on triggered tasks (\S\ref{sec:method_verification}). Table~\ref{tab:trigger_rate} reports the trigger rate, which stays high on most systems but drops on AutoGen, whose short call chains often finish before reaching the fault's injection position. While this degradation is expected, its patterns are not, as shown below. Mini-SE shows the smallest $\Delta$pass@1 because its pass@1 without fault injection is already low (3.47\% to 11.3\% across models), leaving limited room for further degradation. Some cells show negative $\Delta$pass@1 (e.g., AutoGen $-$2.24\% on MMLU-Pro under \llmname{DeepSeek-V3.2}, Mini-SE $-$2.71\% on SWE-bench Pro), because each fault configuration applies to only a few triggered tasks (e.g., 300 / 65 $\approx$ 4.6 per configuration) and LLM outputs vary across independent runs at temperature 0.7. We repeat the full evaluation three times with independent seeds and find the robustness ranking stays similar, so the temperature-induced variation has little effect on our conclusion. 

\begin{table}[t]
\centering
\caption{Average LLM and tool call counts per task without fault injection (w/o) and with fault injection (w/), and their ratio. Backbone LLM is \llmname{Claude-Sonnet-4.5}.}
\label{tab:overhead}
\vspace{-7pt}
\resizebox{0.9\columnwidth}{!}{
\begin{tabular}{l|ccc|ccc}
\Xhline{1.2pt}
\rowcolor{gray!12}
& \multicolumn{3}{c|}{\textbf{LLM Calls}} & \multicolumn{3}{c}{\textbf{Tool Calls}} \\
\rowcolor{gray!12}
\multirow{-2}{*}{\textbf{System}}
& \textbf{w/o} & \textbf{w/} & \textbf{Ratio}
& \textbf{w/o} & \textbf{w/} & \textbf{Ratio} \\
\Xhline{1.2pt}
AutoGen  & 1.72 & 7.3 & 4.24$\times$ & 0.47 & 3.07 & 6.55$\times$ \\
MAD      & 11.74 & 13.32 & 1.14$\times$ & 7 & 7.02 & 1$\times$ \\
MapCoder & 7.34 & 5.2 & 0.71$\times$ & 1.29 & 1.2 & 0.93$\times$ \\
EvoMAC   & 10.76 & 9.31 & 0.87$\times$ & 3.95 & 3.58 & 0.91$\times$ \\
Mini-SE  & 21.73 & 36.07 & 1.66$\times$ & 20.2 & 29.02 & 1.44$\times$ \\
\Xhline{1.2pt}
\end{tabular}
}
\vspace{-12pt}
\end{table}

\begin{table}[t]
\centering
\caption{Trigger rate per system and target field. Backbone LLM is \llmname{Claude-Sonnet-4.5}.}
\label{tab:trigger_rate}
\vspace{-7pt}
\resizebox{0.85\columnwidth}{!}{
\begin{tabular}{lcccc}
\Xhline{1.2pt}
\rowcolor{gray!12}
\textbf{System} & \textbf{Overall} & \textbf{Content} & \textbf{Tool\_calls} & \textbf{Compound} \\
\Xhline{1.2pt}
AutoGen  & 48.30\% & 69.63\%  & 12.41\%  & 69.01\%  \\
MAD      & 99.49\% & 100.00\% & 98.79\%  & 99.42\%  \\
MapCoder & 86.03\% & 100.00\% & 63.79\%  & 95.32\%  \\
EvoMAC   & 98.65\% & 100.00\% & 96.38\%  & 100.00\% \\
Mini-SE  & 87.67\% & 76.73\%  & 100.00\% & 100.00\% \\
\Xhline{1.2pt}
\end{tabular}
}
\vspace{-7pt}
\end{table}

\begin{table}[t]
\centering
\caption{Failure mode breakdown per agent system. Backbone LLM is \llmname{Claude-Sonnet-4.5}.}
\label{tab:failure_mode}
\vspace{-7pt}
\resizebox{0.75\columnwidth}{!}{
\begin{tabular}{lccc}
\Xhline{1.2pt}
\rowcolor{gray!12}
\textbf{System} & \textbf{Aborted} & \textbf{Reported} & \textbf{Silent} \\
\Xhline{1.2pt}
AutoGen  & 1.44\%  & 32.85\% & 65.71\% \\
MAD      & 8.70\%  & 61.12\% & 30.19\% \\
MapCoder & 4.69\%  & 73.49\% & 21.82\% \\
EvoMAC   & 11.85\% & 48.29\% & 39.86\% \\
Mini-SE  & 13.81\% & 25.10\% & 61.09\% \\
\Xhline{1.2pt}
\end{tabular}
}
\vspace{-14pt}
\end{table}

\textbf{Degradation by system implementation.}
Degradation depends on how each system propagates or recovers from faults.
MapCoder is the least robust, dropping 48.61\% on HumanEval, 41.07\% on MBPP, and 38.25\% on MMLU-Pro, because its pipeline feeds each agent's output into the next stage, so a fault propagates through all downstream stages without recovery.
MAD drops 24.2\% on HumanEval and 20.7\% on MATH-500, because a fault on its single moderator propagates directly to the final answer despite aggregation across debaters.
EvoMAC drops 18.48\% on HumanEval and 15.85\% on MATH-500, because its iterative refinement allows later generations to recover from earlier faults, making it the most robust system in our evaluation.
AutoGen drops 19.44\% on HumanEval on triggered tasks, but its low LLM call count ($\mu{=}$1.72 per task) means most configured faults do not trigger, so its effective degradation across all tasks is lower than EvoMAC.
Mini-SE drops only 0.87\% on SWE-bench Pro, because its baseline pass@1 is already low (3.47\% to 11.3\%) and its high LLM call count ($\mu{=}$21.73) dilutes the impact of a single fault.

\textbf{Consistency across models and task domains.}
The degradation pattern is consistent across models. MapCoder, for example, shows the highest $\Delta$pass@1 under every model on every dataset. Our faults modify the LLM response in the same way for all models, so the resulting degradation appears to depend on how the system implementation handles the faulty response rather than on which model produced it. Across task domains, all systems show higher $\Delta$pass@1 on code generation tasks (HumanEval, HumanEval+, MBPP) than on reasoning tasks (MMLU-Pro, MATH-500). For example, under \llmname{Claude-Sonnet-4.5}, MapCoder drops 48.61\% on HumanEval but 38.25\% on MMLU-Pro. Faults on generated code directly produce incorrect programs that fail test cases, while faults on reasoning text may still yield a correct answer if the core logic remains complete.

\textbf{Resource overhead.}
Faults affect LLM and tool call counts per task (Table~\ref{tab:overhead}). AutoGen increases from 1.72 to 7.3 LLM calls (4.24$\times$) and from 0.47 to 3.07 tool calls (6.55$\times$), because faulty responses trigger repeated retries. MapCoder decreases from 7.34 to 5.2 LLM calls (0.71$\times$), because faults cause its pipeline to terminate before reaching later stages. Mini-SE increases from 21.73 to 36.07 LLM calls (1.66$\times$), because faults trigger additional debugging steps. The direction depends on each system's fault handling. Systems that retry after faults consume more, while systems that fail early consume fewer. The wrapper adds less than 1ms per intercepted call, which is negligible compared to LLM API response times that range from 1s to 30s.

\textbf{Failure mode breakdown.}
Beyond the size of the drop, we also examine how each task fails, since an error shown to the user is a safe reaction while a wrong answer with no warning is not. We classify each failed task into Aborted (the run stops), Reported (the output tells the user something went wrong), and Silent (plausible but wrong output with no warning). Table~\ref{tab:failure_mode} reports the breakdown per system. Reported is a safe reaction and covers most failures on MAD and MapCoder, and Aborted runs stay low across all systems, while the share of Silent failures varies widely, from 21.82\% on MapCoder to 65.71\% on AutoGen.

\vspace{-5pt}
\fdbox{
\textbf{Answer to RQ1:} All agent systems degrade under LLM API faults, reaching $\Delta$pass@1 of 49.66\% (MapCoder). The robustness ranking is MapCoder (highest drop), MAD, EvoMAC, AutoGen (lowest drop), and this ranking is similar across all backbone LLMs. Mini-SE shows only 0.87\% drop because its pass@1 without fault injection is already below 12\% and each task makes over 20 LLM calls, so one faulty call has limited effect. Faults also change resource consumption unpredictably (0.71$\times$ to 4.24$\times$).
}

\subsection{RQ2: Impact of Fault Configurations}\label{sec:rq2}

\begin{table}[t]
\centering
\caption{$\Delta$pass@1 (\%) by fault configuration and system. Backbone LLM is \llmname{Claude-Sonnet-4.5}. Results aggregated over all datasets. \textbf{Bold}: highest drop per system within each section.}
\label{tab:config}
\vspace{-7pt}
\resizebox{\columnwidth}{!}{
\begin{tabular}{llccccc}
\Xhline{1.2pt}
\rowcolor{gray!12}
\textbf{Config} & \textbf{Target} & \textbf{AutoGen} & \textbf{MAD} & \textbf{MapCoder} & \textbf{EvoMAC} & \textbf{Mini-SE} \\
\Xhline{1.2pt}
\rowcolor{gray!5}
\multicolumn{7}{c}{\textit{\textbf{Fault Type}}} \\
\hline
Error     & Content   & 23.75 & 37.5 & 59.62 & 42.31 & 0 \\
Error     & Tool call & $-$12.5 & 0 & 0 & 4.05 & 0 \\
Timeout   & Content   & \textbf{28.21} & 22.33 & \textbf{64.42} & \textbf{45.19} & $-$20 \\
Empty     & Content   & 22.5 & \textbf{38.46} & 56.73 & 44.23 & 9.09 \\
Truncate  & Content   & 21.25 & 28.85 & 58.65 & 40.38 & \textbf{14.29} \\
Truncate  & Tool call & 0 & 2.74 & 52.63 & $-$3.95 & 0 \\
Corrupt   & Content   & $-$2.38 & $-$1.92 & 2.88 & 0 & 5.88 \\
Schema    & Content   & 21.69 & \textbf{38.46} & 57.69 & 43.27 & $-$30 \\
Schema    & Tool call & 11.11 & 0 & $-$2.63 & 4.76 & 7.69 \\
\Xhline{0.8pt}
\rowcolor{gray!5}
\multicolumn{7}{c}{\textit{\textbf{Injection Strategy}}} \\
\hline
Single       & & 1.66 & 22.33 & 48.23 & 3.64 & 3.12 \\
Burst        & & 0 & 13.33 & 46.81 & 34.78 & $-$1.69 \\
Intermittent & & $-$3.57 & 8.47 & $-$1.27 & 1.77 & 8.7 \\
Persistent   & & \textbf{57.41} & \textbf{54.74} & \textbf{62.39} & \textbf{47.64} & \textbf{10} \\
\Xhline{0.8pt}
\rowcolor{gray!5}
\multicolumn{7}{c}{\textit{\textbf{Injection Position}}} \\
\hline
1st call & & $-$4.84 & $-$8.06 & \textbf{83.87} & \textbf{6.45} & 0 \\
2nd call & & 11.11 & \textbf{24.19} & 1.61 & $-$1.61 & 0 \\
3rd call & & \textbf{16.67} & 4.84 & $-$4.84 & 3.23 & 0 \\
\Xhline{0.8pt}
\rowcolor{gray!5}
\multicolumn{7}{c}{\textit{\textbf{Compound Scenario}}} \\
\hline
\textit{Single-fault avg.} & & 17.15 & 22.98 & 43.1 & 20.68 & 3.01 \\
\cmidrule(lr){1-7}
API degradation & & 0 & 68.18 & 81.82 & 4.55 & 0 \\
Content filter  & & 0 & 45.45 & \textbf{86.36} & 0 & $-$33.33 \\
Max tokens      & & $-$18.18 & 9.09 & \textbf{86.36} & $-$9.09 & 0 \\
Proxy HTML      & & $-$4.76 & \textbf{76.19} & 80.95 & 0 & $-$25 \\
Slow response   & & 9.52 & $-$9.52 & 9.52 & \textbf{4.76} & 0 \\
Stale cache     & & -- & 0 & 0 & $-$4.76 & 0 \\
Stale data      & & $-$25 & $-$14.29 & $-$5.88 & $-$9.52 & 0 \\
Wrong entity    & & \textbf{33.33} & $-$10 & $-$5.88 & 0 & $-$25 \\
\Xhline{1.2pt}
\end{tabular}
}
\vspace{-14pt}
\end{table}

Beyond overall degradation, we analyze how degradation varies across fault configurations. Table~\ref{tab:config} presents $\Delta$pass@1 by fault type, injection strategy, injection position, and compound scenario, aggregated over all datasets.

\textbf{Fault type and target field.}
Faults targeting the content field cause higher $\Delta$pass@1 than faults targeting the tool call field. Timeout, empty, truncate, schema, and error content faults all cause $\Delta$pass@1 above 15\% on at least one system. The most damaging type differs across systems: timeout for AutoGen (28.21\%), MapCoder (64.42\%), and EvoMAC (45.19\%); empty and schema for MAD (38.46\% each); and truncate for Mini-SE (14.29\%). Among content faults, only corrupt remains below 7\% across all systems ($-$2.38\% to 5.88\%), because agents can recognize and ignore garbled characters. Most tool call faults cause $\Delta$pass@1 below 5\% (e.g., error ranges from $-$12.5\% to 4.05\%), because tool error handling catches them before they propagate to subsequent agents. However, truncate causes 52.63\% on MapCoder because its pipeline passes corrupted tool call arguments to the next stage without validation. Since the most damaging type cannot be predicted in advance, evaluating only obvious fault types risks missing the most harmful one. Mini-SE shows low $\Delta$pass@1 for most types because its baseline pass@1 is already low, although truncate content (14.29\%) and empty content (9.09\%) still cause notable drops. Extreme negative values on Mini-SE (e.g., schema $-$30\%) reflect the few triggered SWE-bench Pro tasks, where one task flip causes large swings.

\textbf{Injection strategy.}
Persistent injection on every LLM call causes the highest $\Delta$pass@1 across systems: MapCoder (62.39\%), AutoGen (57.41\%), MAD (54.74\%), EvoMAC (47.64\%), and Mini-SE (10\%). AutoGen ranks second despite its lowest $\Delta$pass@1 in \S\ref{sec:rq1}, because faults on each call remove its advantage of making few calls per task. Burst injection ranks second for MapCoder (46.81\%) and EvoMAC (34.78\%), as consecutive faults reduce recovery between calls. Single injection causes high $\Delta$pass@1 for MapCoder (48.23\%) and MAD (22.33\%) because a critical fault propagates through downstream pipeline and debate. AutoGen and EvoMAC remain low (1.66\% and 3.64\%) because subsequent rounds recover from a single fault. Intermittent injection has the least impact for most systems because probabilistic triggering produces few injections per task.

\textbf{Injection position.}
Injection-position sensitivity differs across architectures. We inject at the first, second, and third calls because later calls rarely fire on systems with short call chains (Table~\ref{tab:overhead}). Pipeline systems are most sensitive because early faults propagate to all downstream steps. MapCoder ranges from $-$4.84\% to 83.87\% across positions. Debate systems are moderately sensitive, with MAD ranging from $-$8.06\% to 24.19\%. Iterative systems are least sensitive, with EvoMAC ranging from $-$1.61\% to 6.45\%, because later rounds can recover from faults at any position. AutoGen ranges from $-$4.84\% to 16.67\%. Most AutoGen tasks complete in few LLM calls ($\mu{=}$1.72), so its retry mechanism absorbs a fault at the 1st call, and the $-$4.84\% reflects noise from the few triggered tasks.

\textbf{Compound scenario.}
Beyond single faults, compound faults that modify response content cause higher $\Delta$pass@1 than those affecting only latency or metadata. MapCoder shows $\Delta$pass@1 of 86.36\% under max tokens and content filter, 81.82\% under API degradation, and 80.95\% under proxy HTML. MAD shows high $\Delta$pass@1 under proxy HTML (76.19\%) and API degradation (68.18\%). AutoGen shows the highest $\Delta$pass@1 under wrong entity (33.33\%), which replaces named entities in responses. Stale cache shows ``--'' for AutoGen because it replays a previous response, but most AutoGen tasks make one LLM call and have no previous response to replay. Compound faults affecting only latency or metadata cause low $\Delta$pass@1. Slow response, stale cache, and stale data show $\Delta$pass@1 below 10\% for most systems, confirming that latency alone does not affect task correctness. EvoMAC shows low or negative $\Delta$pass@1 across most compound scenarios ($-$9.52\% to 4.76\%), because its iterative refinement recovers from compound faults like single faults. Mini-SE shows 0\% on most compound scenarios because it is evaluated on SWE-bench Pro alone, so each scenario applies to few triggered tasks, and its low baseline pass@1 means most tasks fail regardless of fault injection.

\textbf{Sensitivity to parameters.} We vary the intermittent probability and the burst count on a set of tasks from HumanEval+ and SWE-bench Pro and measure their effect on $\Delta$pass@1 (Table~\ref{tab:sensitivity}). When the probability moves across 10\%, 30\%, and 50\%, each $\Delta$pass@1 changes by less than 5 percentage points, and 30\% falls in the middle of this stable range. The burst count is stable for most systems as well, and does not change beyond a burst of 3.

\begin{table}[t]
\centering
\caption{Sensitivity check on intermittent probability (prob) and burst count (burst). Mini-SE on SWE-bench Pro, others on HumanEval+. Backbone LLM is \llmname{Claude-Sonnet-4.5}.}
\label{tab:sensitivity}
\vspace{-7pt}
\resizebox{\columnwidth}{!}{
\begin{tabular}{lcccccc}
\Xhline{1.2pt}
\rowcolor{gray!12}
\textbf{System} & \textbf{prob=0.1} & \textbf{prob=0.3} & \textbf{prob=0.5} & \textbf{burst=1} & \textbf{burst=3} & \textbf{burst=5} \\
\Xhline{1.2pt}
AutoGen  & 0.00  & 0.00  & 0.00  & 0.00  & 0.00   & 0.00 \\
MAD      & 15.50 & 13.30 & 17.20 & 45.00 & 41.70 & 46.70 \\
MapCoder & 0.00  & 0.00  & 0.00  & 42.40 & 42.40 & 42.40 \\
EvoMAC   & 0.00  & 0.00  & 0.00  & 0.00  & 43.10 & 41.70 \\
Mini-SE  & 0.00  & 0.00  & 0.00  & 0.00  & 0.00  & 0.00 \\
\Xhline{1.2pt}
\end{tabular}
}
\vspace{-7pt}
\end{table}

\vspace{-5pt}
\fdbox{ \textbf{Answer to RQ2:} Fault impact depends on configuration. Content faults cause higher drops than tool call faults, and the most damaging type varies across systems. Persistent injection causes the highest $\Delta$pass@1 (up to 62.39\%). Injection position matters most for pipeline systems, where a fault at the first stage drops pass@1 by up to 83.87\%. Compound faults that modify content amplify degradation (up to 86.36\%).}

\subsection{RQ3: Fault Diagnosis}\label{sec:rq3}

\begin{table}[t]
\centering
\caption{Fault diagnosis accuracy (\%) on Mini-SE (SWE-bench Pro). We evaluate all failed cases under fault injection across all backbone models. Rule: rule-based diagnosis. LLM: LLM-based diagnosis (\llmname{Claude-Sonnet-4.5}). $\kappa$: Cohen's kappa between the two methods.}
\label{tab:fault_diagnosis}
\vspace{-7pt}
\resizebox{0.9\columnwidth}{!}{
\begin{tabular}{lcccccc}
\Xhline{1.2pt}
\rowcolor{gray!12}
& \multicolumn{3}{c}{\textbf{Type Accuracy}} & \multicolumn{3}{c}{\textbf{Step Accuracy}} \\
\rowcolor{gray!12}
\textbf{Fault Type} & \textbf{Rule} & \textbf{LLM} & \textbf{$\kappa$} & \textbf{Rule} & \textbf{LLM} & \textbf{$\kappa$} \\
\Xhline{1.2pt}
Error    & \textbf{47.57} & \textbf{47.57} & 0.813 & \textbf{69.90} & 66.02 & 0.862 \\
Timeout  & \textbf{91.04} & 87.31 & 0.411 & \textbf{89.55} & 88.81 & 0.907 \\
Empty    & \textbf{96.74} & 77.17 & 0.098 & \textbf{38.04} & 31.52 & 0.629 \\
Truncate & 4.3 & \textbf{34.41} & 0.203 & 31.18 & \textbf{40.86} & 0.381 \\
Corrupt  & \textbf{13.64} & 6.36 & 0.163 & \textbf{26.36} & 20 & 0.322 \\
Schema   & \textbf{52.46} & 27.05 & 0.372 & \textbf{63.93} & 60.66 & 0.762 \\
\hline
Overall  & \textbf{52.45} & 47.25 & 0.567 & \textbf{55.5} & 53.52 & 0.661 \\
\Xhline{1.2pt}
\end{tabular}
}
\vspace{-14pt}
\end{table}

We evaluate how well existing methods diagnose the injected fault type and step from execution traces. We focus on Mini-SE on SWE-bench Pro, because Mini-SE produces the longest traces among all tested systems ($\mu{=}$21.73 LLM calls per task), making fault diagnosis most challenging. We select this setting because it represents the hardest case for diagnosis. If methods already struggle on long traces, there is even more reason to improve diagnosis before applying it to production systems.
We run Mini-SE with all backbone models under fault injection and collect all failed cases (654 in total across all models and fault configurations). We apply two diagnosis methods to each case. (1)~\textit{Rule-based diagnosis} is a baseline we design for this evaluation. It scans each step in the execution trace and matches the LLM API response against common fault signatures in order, such as HTTP error codes, timeout messages, empty response fields, encoding corruption patterns, and incomplete text. It returns the first match as the predicted fault type and step. These patterns match general LLM API error formats that developers encounter in production~\cite{26AzureContentFilter}. We write them from public API error documentation rather than from the content we inject, so they are not tailored to our injection implementation.
(2)~\textit{LLM-based diagnosis}~\cite{25ZhangWhich} feeds the full execution trace, the task query, and the system architecture into an LLM in a single prompt and asks it to predict the fault type and the step where the fault first occurred. The prompt lists the fault-type definitions, but does not tell the model which fault we injected or at which step. This method can detect faults that lack fixed patterns, such as truncated or corrupted content, but may struggle to locate the exact step in long traces. Both methods receive the same trace information.

\textbf{Overall diagnosis accuracy.} As shown in Table~\ref{tab:fault_diagnosis}, rule-based diagnosis achieves 52.45\% type accuracy and 55.5\% step accuracy. LLM-based diagnosis achieves 47.25\% type accuracy and 53.52\% step accuracy. Rule-based diagnosis achieves 5.2\% higher type accuracy than LLM-based diagnosis, while both methods achieve comparable step accuracy (within 2\%). Neither method exceeds 56\% on any metric, indicating that current diagnosis methods cannot reliably identify fault type or step from agent execution traces.

\textbf{Accuracy by fault type.} Diagnosis accuracy varies by fault category.
On crash faults, both methods achieve high accuracy. For type accuracy, rule-based diagnosis achieves 91.04\% on timeout and 47.57\% on error, and LLM-based diagnosis achieves 87.31\% and 47.57\%, because crash faults produce distinctive patterns such as HTTP status codes and connection failures. For step accuracy, both methods remain high on timeout (rule-based 89.55\%, LLM-based 88.81\%) and error (rule-based 69.9\%, LLM-based 66.02\%).
On omission faults, the two methods diverge. For empty, rule-based diagnosis achieves 96.74\% type accuracy, while LLM-based diagnosis achieves 77.17\%, because empty content fields are easy to match by fixed patterns but LLM-based diagnosis can confuse them with normal short responses. For truncate, rule-based diagnosis achieves only 4.3\% type accuracy, while LLM-based diagnosis achieves 34.41\%, because truncated output does not match any fixed pattern but LLM-based diagnosis can detect incomplete content. Step accuracy is low for both methods on empty (rule-based 38.04\%, LLM-based 31.52\%) and truncate (rule-based 31.18\%, LLM-based 40.86\%), because omission faults do not leave clear markers at the exact fault step.
On value faults, the two methods both achieve low accuracy on corrupt (rule-based 13.64\% type, LLM-based 6.36\%), because corrupted encoding resembles normal text. Schema faults are easier to detect within this category, with rule-based achieving 52.46\% type accuracy and LLM-based achieving 27.05\%, because schema violations produce recognizable structural errors. Step accuracy is moderate on schema (rule-based 63.93\%, LLM-based 60.66\%) but low on corrupt (rule-based 26.36\%, LLM-based 20\%).

\textbf{Complementarity.} Cohen's $\kappa$ between the two methods is 0.567 for type and 0.661 for step overall, indicating moderate agreement. On crash faults, $\kappa$ is high ($\kappa{=}$0.813 for error type, $\kappa{=}$0.907 for timeout step), because crash faults produce clear signals that both methods can detect. On omission and value faults, $\kappa$ is low ($\kappa{=}$0.098 for empty type, $\kappa{=}$0.203 for truncate type), because rule-based diagnosis relies on structural patterns while LLM-based diagnosis relies on semantic content. Rule-based diagnosis captures empty faults that LLM-based diagnosis misses, while LLM-based diagnosis captures truncate faults that rule-based diagnosis misses. Combining both approaches, for example by using rule-based diagnosis for crash and empty faults and LLM-based diagnosis for truncate faults, can improve overall accuracy.

\vspace{-5pt}
\fdbox{ \textbf{Answer to RQ3:} Neither rule-based nor LLM-based diagnosis exceeds 56\% overall accuracy. Crash faults are easy to diagnose (up to 91.04\%), but omission faults such as truncation, which cause significant damage on most systems (\S\ref{sec:rq2}), are among the hardest to diagnose (4.3\% type accuracy by rule-based, 34.41\% by LLM-based). Value faults such as corruption are also hard to diagnose (6.36\% type accuracy) and hard to detect at runtime. The two methods are complementary ($\kappa{=}$0.567), so routing each fault category to the better method can improve accuracy. }

\section{Discussion}\label{sec:discussion}

\subsection{Implications for Practice}

Our results provide practical guidance for agent system developers and framework designers. We use two terms throughout this section. A fault is \textit{severe} if it causes the response to return a clear, visible error, such as a server error or a timeout~\cite{04AvizienisBasic}. A fault is \textit{harmful} if it causes a large drop in task success ($\Delta$pass@1), regardless of how visible it is~\cite{25HuangAutoInject}. These two properties are independent: a fault can look alarming yet cause little degradation, or look normal yet degrade the task success rate sharply.

\textbf{The most severe faults are not the most harmful.}
Omission faults such as truncation and empty responses cause performance drops comparable to crash faults such as error and timeout on most systems (\S\ref{sec:rq2}), despite appearing less severe. For example, on MAD, empty content causes a 38.46\% drop, close to error content (37.5\%) and higher than timeout content (22.33\%). Crash faults trigger errors and automatic retries, while omission faults return valid HTTP 200 responses and bypass error handling. Developers should add output validation after each LLM API call, such as checking \texttt{finish\_reason}, verifying code syntax completeness, and confirming that tool call arguments match the expected schema.

\textbf{The most harmful faults are also the hardest to diagnose.}
Omission faults such as truncation and empty responses cause large performance drops across most systems yet remain hard to diagnose. Rule-based diagnosis identifies truncation with only 4.3\% accuracy (\S\ref{sec:rq3}). Truncated output looks like weak model output in execution traces, so developers may misattribute the failure to model capability and upgrade the model instead of fixing fault handling. Agent frameworks should log structured metadata for each LLM API call, including token usage relative to the limit, \texttt{finish\_reason}, and response length, to make truncation detectable during later analysis.

\textbf{Robustness depends on system implementation.}
The robustness ranking across systems is similar under all backbone LLMs (\S\ref{sec:rq1}). The reason appears to be that fault handling depends on how the system processes faulty responses rather than on which model produced them. The pipeline system we evaluate (MapCoder) is the most vulnerable: a single fault at its first stage drops pass@1 by up to 83.87\% (\S\ref{sec:rq2}), because each stage consumes the previous output and propagates the fault to all downstream stages. The iterative system is the most robust, as later rounds can observe and correct errors from earlier ones. These mechanisms are tied to how each system structures its LLM calls, so we expect them to carry over to other systems of the same pattern, though confirming this requires evaluating more systems per pattern (\S\ref{sec:Threats to Validity}). Replacing the model alone is unlikely to fix these weaknesses. Developers should add stage-level output validation to pipeline systems and consider iterative refinement to recover from faults.

\subsection{Threats to Validity}\label{sec:Threats to Validity}

For internal validity, we reimplement all agent systems on Google ADK with unified tool interfaces. Although we preserve each system's original interaction logic, behavioral differences from the original implementations may affect absolute pass@1 values. Because the same reimplementation runs with and without fault injection, $\Delta$pass@1 remains a valid robustness measure. We uniformly distribute 65 fault configurations across tasks, so each configuration applies to only a few triggered tasks (e.g., 300 / 65 $\approx$ 4.6 per configuration). This causes $\Delta$pass@1 to fluctuate, as shown by occasional negative values in Table~\ref{tab:overall}. Temperature 0.7 also introduces variance across runs. We reduce this variance by computing $\Delta$pass@1 between two runs on the same task set with deterministic fault injection.

For external validity, our evaluation spans five architectural patterns, seven benchmarks across four task domains, and four backbone LLMs from different providers. Two limitations remain. First, new architectures (e.g., RAG-based agents) may exhibit different fault propagation patterns and are not covered. Second, and more importantly, we evaluate only one agent system per architectural pattern (e.g., MapCoder stands for the pipeline pattern), so within each pattern we cannot separate the effect of the pattern from that of the specific system, which is a significant threat to validity. We therefore read our robustness ranking as a pattern that holds consistently across the systems and models we test, not as a proven property of the architectural patterns themselves. Establishing the latter would require evaluating additional agent systems within each pattern. For construct validity, we measure robustness using $\Delta$pass@1, which captures overall performance degradation but not partial correctness or output quality. We adopt pass@1 because it is the standard metric across our benchmarks. This supports comparison with existing studies and makes the threat minimal. Future work can complement $\Delta$pass@1 with finer-grained metrics such as partial test case pass rates or output similarity scores.

\section{Related Work}\label{sec:relatedwork}
\subsection{Fault Injection for Agent Systems}
Existing fault injection methods can be grouped into agent-oriented offline methods, agent-oriented runtime methods, and infrastructure tools. AgenTracer~\cite{25ZhangAgenTracer} and AutoInject~\cite{25HuangAutoInject} operate offline. AgenTracer uses an LLM to perturb completed execution traces and replays them to construct fault datasets. AutoInject analyzes how fault propagation patterns depend on organizational structures at the trace level. Because these methods operate on static traces rather than running systems, they cannot capture runtime behaviors such as retries or early termination. MAS-FIRE~\cite{26JiaMASFIRE} operates at runtime but is intrusive. It modifies prompts, rewrites responses, and routes messages to inject semantic faults such as hallucination and role ambiguity. However, it requires adapting the injection logic to each system and does not target LLM API faults such as response truncation or encoding corruption. Chaos engineering tools inject infrastructure faults but cannot modify fields within LLM API response bodies. Rainmaker~\cite{23ChenRainmaker} injects transient HTTP faults including 503 errors and timeouts at the REST layer. CrashFuzz~\cite{23GaoCrashFuzz} performs coverage-guided crash and reboot injection for cloud systems. ChaosBlade~\cite{26ChaosBlade} operates at the OS, container, or network layer. These tools can cause API requests to fail, but cannot truncate \texttt{message.content} or corrupt \texttt{tool\_calls} arguments.

In summary, existing methods cannot provide runtime, non-intrusive injection that modifies LLM API response content. Offline methods~\cite{25ZhangAgenTracer, 25HuangAutoInject} cannot inject faults into running systems. Runtime methods~\cite{26JiaMASFIRE} are intrusive and target semantic rather than LLM API faults. Traditional tools~\cite{23ChenRainmaker, 23GaoCrashFuzz, 26ChaosBlade} cannot modify response content. We provide this capability at the LLM API layer. This allows us to inject field-level faults such as truncation and corruption across agent frameworks without modifying their source code.

\subsection{Failure Analysis of Agent Systems}
Existing studies on agent system reliability analyze failure traces after execution through failure mode identification~\cite{25CemriMAST, 26YuFailure}, anomaly detection~\cite{25SolomonLumiMAS}, fault localization~\cite{25ZhangWhich, 25ZhuAgentDebug, 25GeFAMAS}, and error classification~\cite{25DeshpandeTRAIL, 25ZhuRAFFLES}. Empirical studies on agent workflow frameworks~\cite{25XueAgentWorkflowBugs}, AI coding tools~\cite{26ZhangEngineering}, LLM-integrated applications~\cite{25ShaoHydrangea, 26ShaoComfrey, 26TanLIDL, Tan2026LLMRCA}, and open-source LLM deployments~\cite{26YuWhyDoes} further show that failures at LLM API and integration boundaries are common and diverse. However, these studies are observation-based. They neither control which fault occurs or when nor inject faults into running agent systems, so they cannot isolate the effect of a specific LLM API fault type from causes such as model capability limitations or measure its impact on task completion. Security studies address a different problem. AgentFuzz~\cite{25LiuAgentFuzz} and TaintP2X~\cite{26HeTaintP2X} detect injection vulnerabilities through fuzzing and taint analysis, but they focus on malicious inputs rather than infrastructure faults.

\section{Conclusion and Future Work}

We present \ourmethod, a chaos engineering framework for controlled, runtime, non-intrusive LLM API fault injection in agent systems. It defines crash, omission, and value faults on content and tool call fields, injects them through an HTTP layer wrapper without modifying source code, and verifies whether each fault is triggered to filter untriggered tasks. Across agent systems, benchmarks, and backbone models, all systems degrade under fault injection, with $\Delta$pass@1 ranging from 0.87\% (Mini-SE) to 49.66\% (MapCoder). The robustness ranking is consistent across models, suggesting that robustness depends on system implementation. Existing fault diagnosis methods remain below 56\% accuracy, leaving room for improvement. Future work will extend the fault taxonomy and diagnosis methods to cover faults that are hard to detect. We will also design and evaluate defense strategies that validate LLM API responses before downstream use and retry incomplete responses.

\section{Data Availability}
The data and implementation are available on Zenodo~\cite{artifact} and GitHub at \url{https://github.com/IntelligentDDS/AgentChaos}.


\begin{acks}
This work was supported by the National Key Research and Development Program of China (Grant Number: 2024YFB4505904). The corresponding author is Pengfei Chen. This research is supported by the National Research Foundation, under its Investigatorship Grant (NRF-NRFI08-2022-0002). Any opinions, findings and conclusions or recommendations expressed in this material are those of the author(s) and do not reflect the views of National Research Foundation, Singapore.
\end{acks}

\balance
\bibliographystyle{ACM-Reference-Format}
\bibliography{references}


\end{document}